\documentstyle[seceq,mbf,preprint,amssymb,amsopn]{ptptex}

\newcommand{\eqref}[1]{(\ref{#1})}

\newcommand{\ii}{{\mathrm{i}}}
\newcommand{\dd}{{\mathrm{d}}}

\newcommand{\phiy}[2]{
\left({\left(\frac{\de \phi_x}{\de y}\right)^{-1}}\right)^{#1}_{#2}}

\newcommand{\sfeta}{\mib{\eta}}
\newcommand{\sfphi}{\mib{\phi}}
\newcommand{\sfmu}{\mib{\mu}}

\newcommand{\id}{\mathit{id}}

\newtheorem{Thm}{Theorem}[section]

\newtheorem{Rem}[Thm]{Remark}

\newcommand{\braket}[2]{\left\langle{\,{#1}\,,\,{#2}\,}\right\rangle}

\newcommand{\Lie}[2]{{\left[{\,{#1}\,,\,{#2}\,}\right]}}

\newcommand{\Poiss}[2]{\left\{{\,{#1}\,,\,{#2}\,}\right\}}

\newcommand{\bbR}{{\mathbb{R}}}

\newcommand{\de}{\partial}

\newcommand{\calA}{\mathcal{A}}
\newcommand{\calB}{{\mathcal{B}}}

\newcommand{\calO}{{\mathcal{O}}}

\newcommand{\calE}{{\mathcal{E}}}
\newcommand{\calD}{{\mathcal{D}}}
\newcommand{\calX}{{\mathcal{X}}}

\newcommand{\sfA}{{\mathsf{A}}}
\newcommand{\sfB}{{\mathsf{B}}}

\newcommand{\sfS}{{\mathsf{S}}}

\newcommand{\sfX}{{\mathsf{X}}}

\title{On the Globalization of Kontsevich's Star Product and 
the Perturbative Poisson Sigma Model}

\author{Alberto S.~{\sc Cattaneo}$^{1,2,}$\footnote{Work
partially supported by SNF Grant
No.~20-63821.00. E-mail address: asc@math.unizh.ch}
and Giovanni 
{\sc Felder}$^{3,}$\footnote{Work
partially supported by SNF Grant
No.~21-65213.01. E-mail address: felder@math.ethz.ch}}

\inst{$^1$Institut f\"ur Mathematik, Universit\"at Z\"urich--Irchel,\\  
Winterthurerstrasse~190, CH-8057 Z\"urich, Switzerland
\\~
\\
$^2$Department of Mathematics, Harvard University,\\
1~Oxford St., Cambridge,
MA~02138, USA
\\~
\\
$^3$D-MATH, ETH-Zentrum, CH-8092 Z\"urich, Switzerland}

\abst{
The globalization of Kontsevich's local formula (resp., the perturbative
expansion of the Poisson sigma model) is described in down-to-earth terms.
}

\begin{document} 
\maketitle

\section{Introduction}
The problem of deformation quantization consists of deforming, in the realm 
of associative algebras,
the pointwise product of smooth functions on a smooth 
manifold $M$ in the direction of a given Poisson bracket (plus the conditions
that the new product is defined in terms of bidifferential operators that
kill constants). In the case when $M$ is $\bbR^n$, Kontsevich\cite{K}
produced a remarkable formula in terms of the Poisson bivector field
$\alpha$ that generates the Poisson bracket. This formula can
also be viewed\cite{CF1} 
as the perturbative expansion of a certain expectation
value in the so-called Poisson sigma model\cite{I,SS} 
with target $(\bbR^n,\alpha)$ and worldsheet a disk.
As the formula transforms in a very complicated way under diffeormorphisms,
it is a nontrivial task to get a global formula for a generic Poisson
manifold $(M,\alpha)$. This has been described in \citen{K} in terms
of formal geometry and made explicit in \citen{CFT} (see also \citen{CFT1/2}).

The first aim of this paper is to present in 
down-to-earth terms
the globalization of Kontsevich's local formula. 
Its second aim
is to (start to) understand this globalization in terms of the Poisson
sigma model. 

Our exposition is based
on Weinstein's approach\cite{W} and
on the results of \citen{CFT} (rather than on Kontsevich's\cite{K}).
Namely, we proceed as follows.
We identify---e.g., by considering the exponential
map $\phi$ for a torsion-free connection---a neighborhood $U$ 
of the zero section of
the tangent bundle $TM$ at each point $x$ with a neighborhood of $x$ in $M$.
This way
we can use Kontsevich's formula fiberwise on the tangent bundle. 
More precisely, we can express the functions $f$ and $g$ to be multiplied
and the Poisson bivector field $\alpha$ as objects living on $T_xM$ and
use Kontsevich's formula to get a new function on $T_xM$. 
As this
has to be repeated for every point $x\in M$, 
the result will be actually
a function $\sigma_{f,g}$
on $U$. Restricting
this function to the zero section yields finally a new function $f\bullet g$
on $M$,
which we may interpret as the product of $f$ and $g$ (see 
subsection~\ref{ssec-assoc}).
This product is a deformation of the pointwise product along the direction
of the Poisson bracket; it is however in general {\em not}\/ associative.
One may however observe that the product would be indeed associative if,
{\em for every pair of functions $f$ and $g$}\/ on $M$,
the corresponding function $\sigma_{f,g}$ on $U$
were the pullback by $\phi$ of a function on $M$;
for in this case the restriction to the zero section would be the inverse
of the pullback by $\phi$, and associativity of the global product
would be an immediate consequence of the associativity of Kontsevich's
product on each fiber.
We may then try to modify Kontsevich's product to an equivalent one 
on each fiber (with possibly different equivalences on different fibers),
so that the above lucky situation actually occurs. We call
``quantization map'' such a family of equivalences.  
Fortunately, it is possible to prove that quantization maps 
exist.\cite{CFT}

Before delineating the proof, we must recall that Kontsevich's formula
actually depends only on the Taylor expansions at zero of the functions to be
multiplied and of the Poisson bivector field. 
In our case, we have to Taylor expand around the zero section
of $TM$ the pullbacks by $\phi$ of global functions. This way we obtain
particular sections of the the jet bundle $E$, see subsection~\ref{ss-Gc}.
(We may think of sections of $E$ as of functions on an infinitely small
neighborhood $tM$ of the zero section of $TM$.)

The proof of the existence of quantization maps is in three steps. 
First, we recall (see subsection~\ref{ss-Gc})
that sections of $E$ corresponding to global functions on $M$
(as pullbacks by $\phi$ to $tM$) are in one-to-one correspondence 
with horizontal sections for a flat connection $D$ on $E$.
Next (see Sect.~\ref{sec-dqjb}), 
we use Kontsevich's formality theorem (reviewed in Sect.~\ref{sec-kf})
to deform $D$
to a new connection $\calD$ that is a derivation for the fiberwise
product and then
use cohomological arguments to show that it is possible
to further deform $\calD$ to a flat connection $\bar{\calD}$ that is
still a derivation. Finally (see Sect.~\ref{sec-dqPm}), 
again by cohomological arguments, we prove
that there is no obstruction in finding an isomorphism of (formal power
series in the deformation parameter of) sections of $E$ that intertwines
between $D$ and $\bar{\calD}$. This isomorphism is precisely the
quantization map we were looking for.

We conclude the paper (see Sect.~\ref{sec-cpePsm}) by analyzing the above
construction in terms of the Poisson sigma model. We observe that
an exponential
map $\phi$ may be used to define a change of coordinates in the functional
integral that, up to problems on the boundary, preserves the functional
measure and the BV bracket. In the new coordinates and ignoring the boundary
problems, the perturbative expansion yields the globally defined 
non-associative $\bullet$-product described above (and in 
subsection~\ref{ssec-assoc}). It would be very interesting to understand
how (and if) the correct treatment of the boundary produces a quantization
map that yields an associative, global star product.

\section{Kontsevich's formula and formality map}\label{sec-kf}
We recall here the definition\cite{K}
of the formality maps $U$
(which can also be regarded\cite{CF1} as expectation values
of the Poisson sigma model).

Given a collection
$\xi_1,\dots,\xi_n$ of multivector fields on $\bbR^d$ of degrees
$k_1,\dots, k_n$,
one defines
the multidifferential operator $U_n(\xi_1,\dots,\xi_n)$,
which acts on $\ell:=2-2n+\sum_{i=1}^n k_i$ functions, as
follows: 
Let $G_{k_1,\dots,k_n;\ell}$ denote the set of graphs with
$\sum_{i=1}^n k_i+\ell$
numbered vertices such that the $j$th vertex for $j\le \sum_{i=1}^n k_i$
emanates exactly $k_j$ arrows (with the condition that no arrow
ends where it begins). Then
\[
U_n(\xi_1,\dots,\xi_n)
:=\sum_{\Gamma\in G_{k_1,\dots,k_n;\ell}}w_\Gamma\,D_\Gamma,
\]
where $D_\Gamma$ is the multidifferential operator obtained by
putting the multivector field $\xi_j$ on the $j$th vertex and
interpreting each arrow as a partial derivative. The weights $w_\Gamma$
are obtained by certain integrals.
We remark that $\ell$ is defined so that $w_\Gamma$ vanishes for
$\Gamma\in G_{k_1,\dots,k_n;r}$ with $r\not=\ell$.
The formality theorem\cite{K} states that the $U$s satisfy certain quadratic
relations (which can be regarded\cite{CF1} as Ward identities for the Poisson
sigma model).

We are interested only in particular cases of the above formulae; viz.,
when all but at most two of the multivector fields are equal to a given
bivector field $\alpha$ and the remaining (zero, one or two) multivector
fields are vector fields (which we denote momentarily by the letters
$\xi$ and $\zeta$). Then we define
\begin{subeqnarray}\label{PAF}
P(\alpha)&=&
\sum_{j=0}^\infty\frac{\epsilon^j}{j!}\,U_j(\alpha,\dots,\alpha),
\slabel{P}\\
A(\xi,\alpha)&=&
\sum_{j=0}^\infty\frac{\epsilon^j}{j!}\,U_{j+1}(\xi,\alpha,\dots,\alpha),
\slabel{A}\\
F(\xi,\zeta,\alpha)&=&
\sum_{j=0}^\infty\frac{\epsilon^j}{j!}\,
U_{j+2}(\xi,\zeta,\alpha,\dots,\alpha),\slabel{F}
\end{subeqnarray}
where $\epsilon$ is a formal parameter. Observe that $P$ is a
bidifferential
operator, $A$ is a differential operator and $F$ is just a
function. 
For the following developments, it is important to notice that
\begin{subeqnarray}\label{PAFe}
&P(\alpha)(f\otimes g) =fg +\epsilon\alpha(\dd f,\dd g) + O(\epsilon^2),
\quad P(\alpha)(1\otimes f)=P(\alpha)(f\otimes1)=f,
\slabel{Pe}\\
&A(\xi,\alpha)=\xi+O(\epsilon),\slabel{Ae}\\
&F(\xi,\zeta,\alpha)=O(\epsilon),\slabel{Fe}
\end{subeqnarray}
as is proved by a direct computation. 

{}From now on, we assume that $\alpha$ is Poisson.
Then the formality theorem ensures that
$f\star g:=P(\alpha)(f\otimes g)$ defines an associative product on
$C^\infty(\bbR^d)[[\epsilon]]$. This star product is a deformation
quantization of the commutative pointwise product along the direction
of the Poisson structure $\alpha$, as follows from \eqref{Pe}.
Consider now a vector field $\xi$ and its flow $\Phi_t$.
Define $f\star_t g:=P(\Phi_{t*}\alpha)(f\otimes g)$.
Then the formality
theorem implies
\begin{equation}\label{AP}
A(\xi,\alpha) f\star g+f\star A(\xi,\alpha) g -A(\xi,\alpha)(f\star g)=
\frac\dd{\dd t}{\big|_{t=0}}(f\star_t g).
\end{equation}
Hence, $A(\xi,\alpha)=\xi+\cdots$ is a
deformation of the Lie derivative compatible with the star product. 
This deformed Lie derivative is not
however a Lie algebra homomorphism and $F$ actually measures its failure,
as can be read from an another identity in the formality theorem.
\begin{Rem}
The usage of the letters $A$ and $F$ is intentional. In fact,
$A(\bullet,\alpha)$ and $F(\bullet,\bullet,\alpha)$ can be thought
of as  a connection $1$-form and its curvature $2$-form, see
Section~\ref{sec-dqjb}.
\end{Rem}

\section{Formal local coordinates}\label{sec-flc}
Kontsevich's formula requires considering the Taylor expansion of
the functions to be multiplied and of the Poisson bivector field. 
Thus, two functions with the same Taylor expansion at a given point are
indistinguishable from the point of view of the star product.
On the other way, Taylor coefficients transform in a complicated way under
coordinate transformations. Our aim in this section is to give a simple
description of how to deal with this, in part following \citen{W}
(for the more general point of view of formal geometry, see \citen{GK},
\citen{BR} and
\citen{B}).

\subsection{Generalized and formal exponential maps}
Given a smooth $d$-dimensional
(paracompact) manifold $M$, we say that a smooth map
$\phi$ from a neighborhood $U$ of the zero section of $TM$ to $M$
(we write $\phi_x(y)$ for the image of $x\in M$, $y\in T_xM\cap U$)
is a \textsf{generalized exponential map} if $\forall x\in M$
\begin{enumerate}
\item $\phi_x(0)=x$, and
\item $\dd\phi_x(0)=\id$.
\end{enumerate}
An example is the actual exponential map of a 
torsion-free linear connection.\footnote{One may give an even more general
definition replacing 2.\ by the condition that
$\phi_x\colon T_xM\cap U\to M$ is a local diffeomorphism (i.e.,
there is a neighborhood of $0$ in $T_xM\cap U$ mapped diffeomorphically
to a neighborhood of $x$ in $M$). Observe however that such a map induces
a generalized exponential map as follows:
Let 
$g$ be the differential of $\phi$ at $y=0$;  
then define 
$\tilde\phi_x(y)=\phi_x(g^{-1}(x)y)$.}

Let $f$ be a smooth function on $M$. Given a generalized exponential
map $\phi$, we can define for every point $x\in M$ the pullback $\phi_x^*f$,
which is a smooth function on $T_xM\cap U$. We are interested
in the Taylor expansion at $y=0$ of $\phi_x^*f(y)$ which we will denote
by $f_\phi(x;y)$. 
The main observation is that this construction yields the same result
for two different
generalized exponential maps $\phi$ and $\varphi$ if $\forall x\in M$
all partial derivatives of $\phi_x$ and $\varphi_x$ at $y=0$
coincide. We will then identify two such maps.
An equivalence class will be called a 
\textsf{formal exponential map}. 
We can write a formal exponential map
$\phi$ as a collection of formal power series
\[
\phi_x(y) = x + \phi^{(1)}_x(y) + \frac12\phi^{(2)}_x(y)+
\frac1{3!}\phi^{(k)}_x(y)+\cdots
\]
that depends smoothly on $x\in M$.
If we now choose local coordinates 
$\{x^i\}|_{i=1,\dots,d}$ around a given point
$x\in M$, we may explicitly write
\[
\phi^i_x(y)=x^i+y^i+\frac12\phi_{x,jk}^i\, y^jy^k+
\frac1{3!}\phi^i_{x,jkl}\,y^jy^ky^l+\cdots.
\]
The transformation rules
for the coefficients above under a change of coordinates
are determined as follows. Let $\psi$ be the diffeomorphism of the chart
that maps the coordinates $\{x^i\}$ to new coordinates
$\{\bar x^{\bar i}\}$. Let 
$\bar y^{\bar i}=\frac{\de \psi^{\bar i}}{\de x^j}\,y^j$ be the usual
transformation law for vectors. 
Then, the new coefficients in
\[
\bar\phi_{\bar x}^{\bar i}(\bar y) :=\bar x^{\bar i} + \bar y^{\bar i} + 
\frac 12\bar\phi_{\bar x,\bar j\bar k}^{\bar i}\,
\bar y^{\bar j}\bar y^{\bar k}
+\cdots
\]
are defined by the relation
\begin{equation}\label{barphi}
\bar\phi_{\bar x}(\bar y)=\Psi(\phi_x(y)).
\end{equation}
Observe that the coefficients 
$\phi^i_{j_1,\dots,j_k}$ do not transform as tensors
(except under  linear coordinate transformations).
It is an easy and useful exercise to check, e.g., that
$\Gamma^i_{jk}(x):=-\phi^i_{x,jk}$ transforms like the Christoffel symbols
of a torsion-free connection. Indeed, one can construct a formal exponential
map $\phi$ starting from $\Gamma$ via the formal geodesic
flow; viz., consider the equations
\[
\ddot\Phi_x^i +\Gamma^i_{jk}(\Phi_x)\dot\Phi^j_x\dot\Phi^k_x=0,
\]
where $\Phi_x(t,y)$ is formal in $y$ and twice differentiable w.r.t.\ 
$t\in[0,1]$. There is a unique solution with initial conditions
$\Phi_x(0,y)=x$ and $\dot\Phi_x(0,y)=y$; in fact, the above equations
amount to linear differential
equations for the coefficients $\Phi^{(k)}_x$ in the
expansion of $\Phi$ w.r.t.\ $y$. Finally, $\phi_x(y):=\Phi_x(1,y)$ 
is the required formal exponential map.
We display the first orders of such a $\phi$ in local coordinates:
\begin{equation}\label{phigamma}
\phi_x^i(y)=x^i+y^i-\frac12\,\Gamma^i_{rs}(x)\,y^ry^s+
\frac1{3!}\left(2\Gamma^i_{jr}(x)\Gamma^j_{st}(x)-
\partial_r\Gamma^i_{st}(x)\right)
\,y^ry^sy^t+\cdots.
\end{equation}

\begin{Rem}
The above construction shows that formal exponential maps exist. Observe
however that there are formal exponential maps more general than those
obtained by geodesic flows. Such more general formal exponential maps
are needed in certain applications; see, e.g., subsections~{\ref{ssec-sympl}} 
and {\ref{ssec-tr}}.
\end{Rem}

\subsection{The Grothendieck connection}\label{ss-Gc}
Let us come back to our original motivation, i.e., the study of
the Taylor expansion $f_\phi$ of the pullback of a smooth function $f$ via
a formal exponential map $\psi$. We write
\[
f_\phi(x;y)=f_\phi^0(x) + (f_\phi^1)_r(x)\,y^r + 
\frac12 (f_\phi^2)_{rs}(x)\,y^ry^s+
\frac1{3!}(f_\phi^3)_{rst}(x)\,y^ry^sy^t+
\cdots.
\]
The first coefficients are easily computed in local coordinates:
\begin{equation}\label{fphi}
f_\phi^0(x)=f(x),\quad
(f_\phi^1)_r(x) = {\de_r f}(x),\quad
(f_\phi^2)_{rs}(x) = {\de_r\de_s f}(x) +
2 {\de_k f}(x)\,\phi^k_{x,rs},\ \dots.
\end{equation}

Observe that $f_\phi$ is a particular example of a section of
the bundle $E\to M$ (the \textsf{jet bundle})
with fiber $\bbR[[y^1,\dots,y^d]]$ (i.e., formal
power series in $y$ with real coefficients) and transition functions
induced from the transition functions of $TM$ (i.e.,
$E$ is the bundle $F(M)\times_{GL(d)}\bbR[[y^1,\dots,y^d]]$ associated
to the frame bundle $F(M)$ of $M$). In other words, the coefficient
$f^k$ is a covariant symmetric tensor of rank $k$.
For instance,
if $\phi$ is determined using the geodesic flow of a torsion-free
connection as in \eqref{phigamma}, we get
\[
(f_\phi^2)_{rs}(x)\,y^ry^s =\nabla_r\de_sf(x)\,y^ry^s,\quad
(f_\phi^3)_{rst}(x)\,y^ry^sy^t =
\nabla_r\nabla_s\de_tf(x)\,y^ry^sy^t,\ \dots.
\]
where $\nabla$ is the covariant derivative associated to the given connection.

Sections of the form $f_\phi$
have the peculiarity that higher order coefficients are determined
by the zeroth order as in (\ref{fphi}).
More precisely, let $\phi$ be a representative of the given formal
exponential
map. Then $f_\phi(x;y)$ is the Taylor expansion of $f(\phi_x(y))$.
So
\[
\frac{\de f_\phi}{\de x^i} =
\displaystyle\frac{\de f}{\de x^j}\,
\frac{\de \phi_x^j}{\de x^i},\qquad
\frac{\de f_\phi}{\de y^i} =
\displaystyle\frac{\de f}{\de x^j}\,
\frac{\de \phi_x^j}{\de y^i}.
\]
By the second condition on generalized exponential maps, we can invert the 
second relation as
\[
\frac{\de f}{\de x^j} =
\phiy kj\,\frac{\de f_\phi}{\de y^k}.
\]
Thus, we get the equation
\[
D_i f_\phi=0,
\]
where $D_i$ is the operator
\begin{equation}\label{D}
D_i= \frac\de{\de x^i} - R^k_i(x;y)\,\frac\de{\de y^k},
\end{equation}
with
\[
R^k_i(x;y) := \phiy kj\,\frac{\de \phi_x^j}{\de x^i}.
\]
If $\xi=\xi^i\frac\de{\de x^i}$ 
is a vector at $x$, we write $D_\xi:=\xi^iD_i$ and,
by (\ref{D}), we have
\begin{equation}\label{Dxi}
D_\xi=\xi+\hat\xi
\end{equation}
with
\begin{equation}\label{hatxi}
\hat\xi^k(x;y)=-\xi^i\,R^k_i(x;y)\,\frac\de{\de y^k}.
\end{equation}
For every $x$, $\hat\xi(x,\bullet)$ is a formal vector field in $y$.
Given a section $\sigma$ of $E$, 
one may easily see that
\[
D_\xi\sigma(x;y)=
{\frac\dd{\dd t}}{\big|_{t=0}}\sigma(x(t);\phi_{x(t)}^{-1}(\phi_x(y))),
\] 
where $x(t)$ is any curve such that $x(0)=x$ and $\dot x(0)=\xi$.
This in particular shows that the definition of $D_\xi$ is independent
of the choice of coordinates. Moreover, by repeatedly applying the above
formula, one sees that $\Lie{D_X}{D_Y}=D_{\Lie XY}$, where $X$ and $Y$
are vector fields on $M$. One may summarize these properties by defining
the covariant derivative
$D:=D_i\,\dd x^i\colon\Gamma(E)\to\Omega^1(M,E)$ and saying that it is flat,
i.e., $D^2=0$.  This is sometimes called the 
\textsf{Grothendieck connection}. 

Observe then that $R^k_i(x;y)$ is a formal power series in $y$ which begins
with $\delta^k_i$ and whose coefficients are smooth
in $x$. By this properties it follows immediately that the coefficients
of a section $\sigma$
of $E$ satisfying $D\sigma=0$ are determined by the zeroth coefficient
$\sigma^0(x)$. If we set $\tilde \sigma =(\sigma^0)_\phi$,
we have $D(\sigma-\tilde\sigma)=0$ and $(\sigma-\tilde\sigma)|_{y=0}$;
but this implies $\sigma=\tilde\sigma$.
In other words,
a section of $E$ is the Taylor expansion of a globally defined function
if{f} it is $D$-closed.

The sections of $E$ form an algebra defined by the Cauchy rule on the
coefficients (extending the product of polynomials in $y$ to formal power 
series). Sections of the form $f_\phi$ clearly form a subalgebra.
Actually more is true; viz.,
$D$ is a derivation (i.e., 
$D(\sigma\tau)=D\sigma\,\tau+\sigma\,D\tau$), so the algebra of
global functions can be
identified with the subalgebra of $D$-closed sections. 

In summary, there is a connection $D$ on the bundle $E$ with the
following properties:
\begin{enumerate}
\item $D$ is a derivation;
\item $D$ is flat; and
\item the subalgebra of $D$-closed sections is isomorphic to the
algebra of smooth functions on $M$.
\end{enumerate}
The aim of the following sections is to deform (``quantize'') the above
properties. In order to do so, we will repeatedly use the fact that
$D$-cohomology is almost trivial.
To see this, it is useful to define the total degree of a form on $M$
taking values in sections of $E$ as the sum of the form degree and the
degree in $y$. Then we write
\[
D=-\delta+D',
\]
where
\[
\delta:=\dd x^i\,R^k_i(x;0)\,\frac\de{\de y^k}=
\dd x^i\,
\frac\de{\de y^i}
\]
is the zero-degree part and $D'$ has positive degree.
It follows immediately that $\delta^2=0$.
We can define a dual operator to $\delta$: viz.,
\begin{equation}\label{delta*}
\delta^*:=y^i\,
\iota_{\frac\de{\de x^i}},
\end{equation}
where $\iota$ denotes contraction with the corresponding vector.
It is easy to verify that $(\delta\delta^*+\delta^*\delta)\rho=k\rho$
for every form $\rho$ of total degree $k$. Thus, if we restrict to
$\delta$-closed 
forms of positive total degree $k$, we may then invert $\delta$ by
$\delta^{-1}\rho=\frac 1k\delta^*\rho$. This inverse yields the unique
form $\sigma$ such that $\delta\sigma=\rho$ and $\delta^*\sigma=0$.
This proves that the $\delta$-cohomology is concentrated in degree zero,
i.e., functions on $M$ (independent of $y$).
By induction, one can prove that the same result holds for $D$.

\subsection{Multivector fields in formal local coordinates}
Given a point $x\in M$, by assumption we can invert the map $\phi_x$.
We can then consider the push-forward $(\phi_x)^{-1}_*F$
of a multivector field $F$ defined on $M$. We will denote by $F_\phi$
its Taylor expansion, which is then, for any $x\in M$, a formal multivector
field in $y$.
For example, if $X$ is a vector field on $M$, then
\[
X_\phi^i(x;y)=X^j(\phi_x(y))\,\phiy ij.
\]
If $\phi$ is determined by a connection as in \eqref{phigamma},
the first orders in the expansion of $X_\phi$ are 
\begin{equation}\label{Xphigamma}
X_\phi^i(x;y)=X^i(x)+
\nabla_rX^i(x)\,y^r+
\left(\frac12\,\nabla_r\nabla_sX^i(x)+
\frac16\,R^i_{rjs}(x)X^j(x)\right)\,y^ry^s+\cdots,
\end{equation}
where $\nabla$ is the covariant derivative associated to the given connection
and $R$ is its curvature tensor.

Observe that the construction (\ref{hatxi})
of a formal vector field in $y$ starting from a vector
$\xi$ at $x$ can geometrically be
understood as follows: Let $\Phi_t$ be the flow of a vector field that
extends $\xi$ in a neighborhood of $x$. We may then consider the
family of formal coordinates that map $0$ to $\Phi_t$
(i.e., $\phi_{\Phi_t}$).
To a point $z$ in
the neighborhood of $x$ we then associate the family
$\phi_{\Phi_t}\circ\phi_x^{-1}$.
Differentiating w.r.t.\ $t$ at $t=0$ gives a family $X$ of vector fields
parametrically depending on $y$ (and independent of how $\xi$ has been
extended). Then $\hat\xi$ is just $-X_\phi$.

We conclude by considering the case of a bivector field $\alpha$.
Its expression $\alpha_\phi$ in formal local coordinates is a bivector
field in $y$ depending smoothly on $x$. 
If $\phi$ is determined by a connection as in \eqref{phigamma},
then by \eqref{Xphigamma} and by the multilinearity of the push-forward,
we get
\begin{equation}\label{alphaphigamma}
\alpha_\phi^{ij}(x;y)=\alpha^{ij}(x)+
\nabla_r\alpha^{ij}(x)\,y^r+
\left(\frac12\,\nabla_r\nabla_s\alpha^{ij}(x)-
\frac16\,R^{[i}_{rks}(x)\alpha^{j]k}(x)\right)\,y^ry^s+\cdots,
\end{equation}
where $[\ ]$ denotes antisymmetrization of the enclosed indices:
$X^{[ij]}:=X^{ij}-X^{ji}$.

If $\alpha$ is Poisson, so is
$\alpha_\phi$ since the push-forward preserves the Lie bracket.
Thus, a Poisson structure on $M$ induces a Poisson structure on sections
of $E$:
\[
\Poiss\sigma\tau(x;y):=\frac12\,\alpha_\phi^{ij}(x;y)\,
\frac{\de\sigma(x;y)}{\de y^i}\,
\frac{\de\tau(x;y)}{\de y^j},
\qquad\sigma,\tau\in\Gamma(E).
\]
We will see in the next section that Kontsevich's formula may be used
to deform the algebra of sections in the direction of the Poisson bracket.

\subsection{Formal geometry}
We would like to end this Section by putting its content in relation with
the language of formal geometry used in \citen{CF1}.
There we chose a section $\phi^{\mathrm aff}$
of the bundle $M^{\mathrm aff}\to M$,
where $M^{\mathrm aff}$ is the quotient by $GL(d)$ of the manifold
$M^{\mathrm coor}$ of jets of coordinate systems.
Given such a section, consider a covering of $M$ by contractible
(or, actually, just parallelizable) open sets.
The restriction of $\phi^{\mathrm aff}$ to an open set $U$ of the covering
may be lifted to a section $\phi_U$
of $U^{\mathrm coor}\to U$, and this
lift is unique if we further assume that the differential of $\phi_U$
w.r.t.\ $y$ at $y=0$ is the the identity in $GL(d)$ for every $x\in U$.
These local expressions
of the formal local coordinates then transform precisely as in \eqref{barphi}.


\section{Deformation quantization of the jet bundle}\label{sec-dqjb}
We have seen in the previous Section that $\Gamma(E)$ is an algebra
(over $C^\infty(\bbR)$) and that, moreover, it is a Poisson algebra
with formal Poisson bivector field $\alpha_\phi$ if $\alpha$ is a Poisson
bivector field on $M$. The construction mentioned in Section~\ref{sec-kf}
applies without modifications to \emph{formal}\/ multivector fields as well.
In particular, $P(\alpha_\phi)$ defines, pointwise w.r.t.\ $x\in M$,
an associative product on sections
of $E$,
\[
\sigma\star\tau:=P(\alpha_\phi)(\sigma\otimes\tau),\qquad
\sigma,\tau\in\Gamma(E)[[\epsilon]],
\]
which is a deformation of the commutative product in the direction of
$\alpha_\phi$ because of \eqref{Pe}.
In the following we will denote by $\calE$ the bundle of $\star$-algebras
whose section are formal power series in $\epsilon$ of sections of $E$ 
(viz., $\calE=F(M)\times_{GL(d)}\bbR[[y_1,\dots,y_d]][[\epsilon]]$).

Our next aim is to find a subalgebra of $\Gamma(\calE)$ that is a
deformation quantization of the subalgebra $C^\infty(M)$ of $\Gamma(E)$.
To do so, we look for a ``quantization" of the Grothendieck connection $D$
and try to define the deformation quantization of $C^\infty(M)$ as the
subalgebra of closed sections. For this program to work, we need the
quantum connection to be a derivation and to be flat.
Given a vector $\xi$ at $T_xM$, our first guess (which will prove
not to be enough) is to define the quantum
covariant derivative in the direction of $\xi$ simply by replacing
$\hat\xi$ in (\ref{Dxi}) by $A(\hat\xi,\alpha_\phi)$; viz., we define
\[
\calD_\xi=\xi+A(\hat\xi,\alpha_\phi)=D_\xi+O(\epsilon).
\]
It can be proved that $\calD$ is well defined globally.
Moreover, (\ref{AP}) implies that $\calD$ is a derivation.
On the other hand, $\calD$ is in general
\emph{not}\/ flat, but at least the formality theorem ensures that 
$\calD^2$ is an inner derivation;
actually, we have
\[
\calD^2\sigma=\Lie{F^M}\sigma_\star:=F^M\star\sigma-
\sigma\star F^M, \qquad\sigma\in\Gamma(\calE),
\]
where $F^M$, called the \textsf{Weyl curvature} of $\calD$,
is the $2$-form on $M$ taking values in sections of $\calE$
defined by
\[
F^M(\xi,\zeta)=F(\hat\xi,\hat\zeta,\alpha_\phi).
\]
Finally, a further identity in the formality theorem
implies the \textsf{Bianchi identity}
\[
\calD F^M=0.
\]
In general, a connection on a bundle of associative algebras with
the above properties (viz., to be a derivation whose curvature
is an inner derivation such that its Weyl curvature satisfies
the Bianchi identity) is called a \textsf{Fedosov connection}.

We want now to modify $\calD$ so that it becomes flat still remaining
a derivation. The first observation is that
\[
\bar{\calD}:=\calD+\Lie\gamma{\ }_\star
\]
is a derivation for any $1$-form $\gamma$ on $M$ taking
values on sections of $\calE$. Moreover, $\bar{\calD}$ turns out to
be again a Fedosov connection. Its Weyl curvature is promptly computed as
\[
\bar F^M=F^M+\calD\,\gamma+\gamma\star\gamma.
\]
If we are now able to find $\gamma$ so that $\bar F^M=0$ (or,
more generally, so that $\bar F^M$ is \textsf{central}, i.e., it commutes
with every section of $\calE$), then $\bar{\calD}$-closed sections will form
a nontrivial subalgebra of $\calE$ (our next step---see 
Sect.~\ref{sec-dqPm}---will then be to prove that
this subalgebra actually provides a deformation quantization of
$C^\infty(M)$).

This program actually works as we will sketch in the following.
Since $F$ (and so $F^M$) starts
at order $\epsilon$, see \eqref{Fe}, we may write
$F^M=\epsilon F_1+\epsilon^2 F_2+\cdots$. 
We write $\calD=D+\epsilon \calD_1+\cdots$ and
$\gamma=\gamma_0+
\epsilon\gamma_1+\epsilon^2\gamma_2+\cdots$. Finally,
we write $\bar F^M=\epsilon\bar F_1+\epsilon^2\bar F_2+\cdots$, and we want
to show that we can set
$\bar F^M=\omega:=\omega_0+\epsilon\omega_1+\cdots$, where $\omega$
is a $2$-form taking values in sections of $\calE$ with the property
that $\Lie\omega{\ }_\star=0$. For example, we may take $\omega=0$.
Observe that the Bianchi identity for $\bar F^M$ implies that
necessarily
$\bar{\calD}\omega=\calD\omega=0$.
The equation $\bar F^M=\omega$ up to order $\epsilon$ reads
\[
D\gamma_0=\omega_0,\qquad F_1+D\gamma_1+\calD_1\gamma_0
+\frac12\{\gamma_0,\gamma_0\}=\omega_1,
\]
while the $\calD$-closedness condition on $\omega$ implies
at this order $D\omega_0=0$. In particular, $D\gamma_0$ is equal to a $D$-closed
expression. But, as observed at the end of subsection~\ref{ss-Gc},
the $D$-cohomology is trivial, so it is possible to find a $\gamma_0$
that solves the equation. Moreover, $\gamma_0$ is uniquely determined
by the ``gauge-fixing'' condition $\delta^*\gamma_0=0$, with $\delta^*$
defined in \eqref{delta*}. The Bianchi identity and
the conditions on $\omega$ imply $DF_1=0$, $\{\omega_0,\gamma_0\}=0$ and $D\omega_1+
\calD_1\omega_0=0$. It follows that also
$D\gamma_1$ is equal to a $D$-closed, and therefore exact, form.
At higher order in $\epsilon$, one proves by induction that one always
has an equation of the form $D\gamma_k$ equal to a $D$-closed form depending
on the lower order coefficients of $\gamma$, $F^M$ and $\omega$.

\section{Deformation quantization of Poisson manifolds}\label{sec-dqPm}
In Section~\ref{sec-flc} we have seen that the algebra of smooth functions
on the manifold $M$ is isomorphic to the subalgebra $A$ of $D$-closed
sections of the jet bundle $E$.
If $M$ is a Poisson manifold, we may deform the algebra of sections
of $E$ in the direction of the Poisson structure in the given formal local
coordinates; we have denoted by $\calE$ the deformed bundle of algebras.
Moreover, as described in Section~\ref{sec-dqjb},
we may define a flat connection $\bar{\calD}$ that is also a derivation
on sections of $\calE$; so we may consider the subalgebra $\calA$
of $\bar\calD$-closed sections. We want to prove that $\calA$ provides
a deformation quantization of $M$, i.e., that there is a module
isomorphism between $A[[\epsilon]]$ and $\calA$ such that
the product on $\calA$ is a deformation of the product on the image of
$A[[\epsilon]]$ in the direction of the Poisson bracket
(plus the usual conditions).
More precisely, we construct a map $\rho$ from formal power series
in $\epsilon$ of sections of $E$ to sections of $\calE$
that deforms the identity map and satisfies
$\bar\calD\rho(\sigma)=\rho(D\sigma)$ for every 
$\sigma\in\Gamma(E)[[\epsilon]]$. It is possible to prove that this map
exists since there  are again no cohomological obstruction. In particular,
it is possible to find $\rho$ of the form 
$\rho={\mathit{id}}+\sum_{k=1}^\infty \epsilon^k \rho_k$,
where $\rho_k$ is a
differential operator w.r.t.\ $y$ of order $\le k$, vanishing on
constants and depending smoothly on $x\in M$. 
For a given $\bar\calD$, 
there is moreover a unique $\rho$ 
satisfying
$\rho|_{y=0}=\mathit{id}$.
Finally, we may define a global star product on $M$ by
\begin{equation}\label{starM}
f\star_M g:=  
\left[\rho^{-1}\left(\rho(f_\phi)\star\rho(g_\phi)\right)\right]|_{y=0}.
\end{equation}

\subsection{Another viewpoint: gaining associativity}\label{ssec-assoc}
Given two smooth functions $f$ and $g$ on $M$, one may be 
tempted---as described in the Introduction---to define
a product of the form
\begin{equation}\label{bullet}
f\bullet g(x)= \left\{P(\alpha_\phi)(f_\phi\otimes g_\phi)\right\}(x;0).
\end{equation}
This product is well-defined on $C^\infty(M)[[\epsilon]]$; however, it is
not associative in general. It would be if the maps
\[
L\colon 
\begin{array}[t]{ccc}
C^\infty(M) &\to& \Gamma(E)\\
f &\rightsquigarrow& f_\phi
\end{array}
\quad{\mathrm{and}}\quad
G\colon 
\begin{array}[t]{ccc}
\Gamma(E) &\to& C^\infty(M)\\
\sigma &\rightsquigarrow& \sigma|_{y=0}
\end{array}
\]
were inverse to each other (whereas $G$ is only a left inverse to $L$).
In this case 
$f\bullet g(x)=G\left(P(\alpha_\phi)(L(f)\otimes L(g))\right)$
would be equivalent on each fiber of $E$ to the associative product
defined by $\alpha_\phi$. 
The way out is to correct the above formula by introducing a 
``quantization map'' $\rho$ with the property that
$\rho^{-1}(P(\alpha_\phi)(\rho(L(f))\otimes \rho(L(g))))$ 
is in the image of $L$
so that $G$ applied to it is actually $L^{-1}$. We finally get the
product 
\[
f\star_Mg = L^{-1}(\rho^{-1}(P(\alpha_\phi)(\rho(L(f))\otimes \rho(L(g))))),
\]
which is the same as \eqref{starM} and is clearly associative as it is now
fiberwise equivalent to the associative product defined by $P(\alpha_\phi)$.
Quantization maps exist as described
in the previous section.

\subsection{The symplectic case}\label{ssec-sympl}
The symplectic case was solved long ago by DeWilde and Lecomte\cite{DL}
and constructively by Fedosov.\cite{Fed}.
The peculiarity of the symplectic case is that locally one may
choose Darboux coordinates, so that the symplectic form and consequently
the Poisson bivector field become constant. At this point one may use
Moyal's star product. The problem is again that of
gluing the local products together.

In the framework of the present paper, this can be reformulated
as follows---and it is closer in spirit to Omori, Maeda and Yoshioka's
approach.\cite{OMY}
First, one chooses
a  formal exponential map $\phi$ with the property that 
$\alpha_\phi(x;y)$ is independent of $y$---so that Kontsevich's product
reduces to Moyal's---and then one looks for a quantization map as
explained above. It must be observed that in general no
formal geodesic flow has this
property, so that one has to consider more general formal exponential maps.
Actually, given any formal exponential map $\phi_0$ we may correct it
by computing the formal diffeomorphism of $T_xM$ that makes
$\alpha_{\phi_0}(x;y)$ constant in $y$. This diffeomorphism can be obtained
using Moser's method observing that $T_xM$ is contractible. 
The construction can be made
smooth in $x\in M$
by choosing globally a way to contract the fibers of $TM$ to the
zero section (e.g., by scaling).
We refer to \citen{CFT1/2,CFT2} for further details.

\subsection{Traces}\label{ssec-tr}
It is proved in \citen{T}, \citen{S}, \citen{TT}
that to every distribution that annihilates
all Poisson brackets there corresponds a trace for the star-product.
We may obtain this result in the framework of this paper in the simpler
case when 
the distribution is given by a volume form $v$.
In this case, essentially by reasoning as in the previous subsection,
one can construct a formal exponential map (usually not coming from
a formal geodesic flow), so that $\phi^*v(x;y)$ is independent of $y$.
The star product defined on the fiber has then a trace\cite{FS} 
provided by integration against $\phi^*v$. If 
\[
\rho(x;y;\epsilon)
:=\sum_k\sum_{i_1,\ldots,i_k}\rho^{i_1,\ldots,i_k}(x;y;\epsilon)
\frac\de{\de y^{i_1}}\cdots\frac\de{\de y^{i_k}}
\]
is the corresponding quantization map, we define the function
\[
\sigma(x)=
\sum_k(-1)^k\sum_{i_1,\ldots,i_k}
\frac\de{\de y^{i_1}}\cdots\frac\de{\de y^{i_k}}
\rho^{i_1,\ldots,i_k}(x;y;\epsilon)\Big|_{y=0}.
\]
Then one can prove---see \citen{CFT2} for details---that
${\rm Tr}\,f:=\int_M f\sigma v$ is a trace 
on compactly supported functions.


\section{Covariant perturbative expansion of the Poisson sigma 
model}\label{sec-cpePsm}
The Kontsevich formula may be obtained\cite{CF1} 
from the perturbative expansion
of the Poisson sigma model,\cite{I,SS} whose action reads
\[
S^{\rm cl}=\int_\Sigma \left(
\braket\eta{\dd X}+\frac12(\alpha\circ X)(\eta,\eta)
\right).
\]
Here $\alpha$ is a given Poisson vector field on a manifold $M$,
$X$ is a map $\Sigma\to M$ (with $\Sigma$ a $2$-dimensional surface, usually
a disk) and $\eta$ is a $1$-form on $\Sigma$ taking values in covectors
at the image of $X$ (i.e., $\eta\in\Gamma(\Sigma,T^*\Sigma\otimes X^*T^*M)$).
Due to the existence of symmetries that do not close off-shell, 
one has to resort to the BV formalism. 
Using the notations of \citen{CF3},
one may organize fields, ghost, and
antifields into superfields $(\sfX,\sfeta)$ which constitute the components
of a supermap $\Pi T\Sigma\to\Pi T^*M$.\footnote{Given a vector bundle $V$,
we denote by $\Pi V$ the supermanifold obtained by reversing the parity of 
its fibers.}
This space of maps can also be regarded as the odd cotangent bundle
$\Pi T^*\calX$ of the space $\calX$ of supermaps $\Pi T\Sigma\to M$ and
has a canonical odd symplectic structure which generates the BV bracket.
If we denote by $\theta$ the odd
coordinates on $\Pi T\Sigma$, we may explicitly write $\sfX\in\calX$ and
$\sfeta\in\Pi T^*_{\sfX}\calX$ as
\begin{eqnarray*}
\sfX &=& X 
+ \theta^\mu\eta^+_\mu-\frac12\theta^\mu\theta^\nu\beta^+_{\mu\nu},\\
\sfeta &=& \beta + \theta^\mu\eta_\mu+\frac12\theta^\mu\theta^\nu X^+_{\mu\nu},
\end{eqnarray*}
where $\beta$ is a ghost (and has ghost number one), while
$\eta^+$, $\beta^+$ and $X^+$ are antifields (of ghost number
$-1$, $-2$ and $-1$ respectively). 
The full BV action then reads
\[
\sfS =  \int_{\Pi T\Sigma}\left(-\braket{D\sfX}\sfeta+
\frac12(\alpha\circ\sfX)(\sfeta,\sfeta)\right)\;\sfmu,
\]
where $\sfmu$ is the canonical supermeasure 
$\dd\theta^2\dd\theta^1\dd u^1\dd u^2$ on $\Pi T\Sigma$ and
$D$ is the canonical cohomological vector field 
$\theta^\mu\frac\de{\de u^\mu}$.

In order to write the BV action more explicitly, one must expand the integrand
in powers of $\theta$ and integrate against $\dd\theta^2\dd\theta^1$,
thus getting an honest differential form to integrate on $\Sigma$.
However, to do this, one needs either to introduce local coordinates---as
is done in \citen{CF1}---at the price of getting a noncovariant action,
or---as we will do now---to pick a formal
exponential map
$\phi$ to perform a field redefinition that allows one to write a covariant
action. Since we are interested here only in describing perturbative
expansions, we will follow the second approach {\em after}\/ choosing 
a critical
point of the action. To simplify the treatment, we restrict ourselves
to the case of interest for deformation quantization: viz., we pick a critical
point of the form $\sfX(u,\theta)=x$, $\sfeta(u,\theta)=0$, 
$\forall(u,\theta)\in\Pi T\Sigma$.
We then denote by $\calX_x$ some neighborhood in the space of 
supermaps $\calX$ of $\sfX\equiv x$. For the perturbative expansion around
$x$, it is then enough to restrict the space of fields to $\Pi T^*\calX_x$.

Next we introduce a space of fields adapted to this critical solution.
Namely, let $\calB^x$ be the space of supermaps $\Pi T\Sigma\to T_xM$, and
let $\calB^x_0$ denote some neighborhood in $\calB^x$ of the zero map.
Now, given a formal exponential map $\phi$, we may define a supermap
$\sfphi_x\colon\calB^x_0\to\calX_x$, $\sfB\rightsquigarrow\sfX$, by
\begin{equation}\label{sfphix}
\sfX(u) = \phi_{x}(\sfB(u)),\qquad\forall u\in\Sigma.
\end{equation}
If we choose the neighborhoods $\calB^x_0$ and $\calX_x$ appropriately,
this map is a diffeomorphism. This map can then be canonically
extended to the odd cotangent bundles by 
$\sfphi_\sfB\colon\Pi T^*_\sfB\calB^x_0\to\Pi T^*_{\sfphi(\sfB)}\calX_x$,
$\sfA\rightsquigarrow\sfeta$:
\begin{equation}\label{sfphiB}
\sfeta(u) = (\dd\phi_{x}(\sfB(u)))^{-1,T}\,\sfA(u),\qquad
\forall u\in\Sigma.
\end{equation}
Since the map $\Pi T^*\calB^x\to\Pi T^*\calX$ defined above is a 
symplectomorphism and (at least formally) unimodular, the perturbative
expansion of the Poisson sigma model with fields $(\sfX,\sfeta)$ around
$\sfX\equiv x$ and $\sfeta\equiv0$ coincides with the perturbative
expansion in the new fields $(\sfB,\sfA)$ around $\sfB\equiv0$, $\sfA\equiv0$.
It is not difficult to compute the action in the new fields: viz., one gets
\[
\sfS =  \int_{\Pi T\Sigma}\left(-\braket{D\sfB}\sfA+
\frac12\alpha_\phi(x;\sfB)(\sfA,\sfA)\right)\;\sfmu.
\]
This can now be expanded in powers of $\theta$ and reduced to an ordinary
integration over $\Sigma$ of a function of the field components
\begin{subeqnarray}\label{sfBA}
\sfB &=& B 
+ \theta^\mu A^+_\mu-\frac12\theta^\mu\theta^\nu c^+_{\mu\nu},\label{sfB}\\
\sfA &=& c + \theta^\mu A_\mu+\frac12\theta^\mu\theta^\nu B^+_{\mu\nu}.
\label{sfA}
\end{subeqnarray}
Observe that all field components are forms on $\Sigma$ taking values
either in $T_xM$ or $T^*_xM$.
The kinetic term of the action may now be expanded as
\[
S_0 = \int_\Sigma A_i\wedge \dd B^i
-A^{+i}\wedge \dd c_i.
\]
The interaction term may then be expanded as in \citen{CF1} in terms
of $\alpha_\phi(x;y)$ and its first two
derivatives w.r.t.\ $y$ evaluated at $y=B$. One may then further
Taylor expand in $B$. The result will then be a sum
of interaction terms in all the fields and antifields with coefficients
given by derivatives w.r.t.\ $y$
of $\alpha_\phi(x;y)$ at $y=0$.
Observe that, since they are all tensors at $x$ (like the fields and the 
antifields), the action obtained this way is covariant.

\subsection{Gauge fixing}
In order to gauge-fix this model, one has first to introduce an antighost
$\bar c$ and a Lagrange multiplier $\mu$, both maps
from $\Sigma$ to $T_xM$, 
the first with ghost number minus one, 
the second with ghost
number zero. Their antifields $\bar c^+$ and $\lambda^+$ are instead
$2$-forms on $\Sigma$ taking values in
$T^*_xM$
with ghost number zero and minus one, respectively.
To the action one adds the term $-\int_\Sigma\braket\mu{\bar c^+}$.
After introducing a gauge-fixing fermion $\Psi$ (a function of the fields
of ghost number minus one), one sets each antifield to be the derivative
(say, from the left) of $\Psi$ w.r.t.\ the corresponding field.
A typical gauge fixing is $\dd{*A}=0$ where $*$ is a Hodge star
operator on $\Sigma$. This corresponds to choosing the gauge-fixing fermion
\begin{equation}\label{Psi}
\Psi=-\int_\Sigma \braket{\dd\bar c}{*A},
\end{equation}
which yields
\[
A^{+i} = *\dd\bar c^i,\qquad
\bar c^+_i = \dd {* A_i},
\]
while $B^+$, $c^+$ and $\mu^+$ are set equal to zero.

\subsection{Deformation quantization}
We assume from now on that $\Sigma$ is a disk and choose boundary
conditions as in \citen{CF1}. To fix the constant in the critical
solution $\sfX\equiv\rm{const.}$,
$\sfeta=0$, we impose
the condition $X(\infty)=x$, where $\infty$ is
a point on the boundary of $\Sigma$. This amounts to the condition
$B(\infty)=0$. Now, given two smooth functions $f$ and $g$ on $M$,
we may compute the expectation value
of $f(X(u))\,g(X(v))$, where $u$ and $v$ are two ordered points on the boundary
of $\Sigma$ minus $\{\infty\}$. 
The perturbative expansion of this expectation value---which, after the
field redefinition, corresponds to the expectation value of
$f_\phi(x;B(u))\,g_\phi(x;B(v))$---can be
computed exactly as in 
\citen{CF1} and yields the function we denoted by
$f\bullet g(x)$
in \eqref{bullet}. 
Compared to Kontsevich's formula as
obtained in \citen{CF1} by first choosing local coordinate and then introducing
the whole BV machinery, the present approach has the advantage of yielding
a well-defined global formula (which depends however, though
in a controlled way, on the choice of $\phi$). The drawback is that the
bullet product is {\em not} associative. 
As we have seen in \ref{ssec-assoc}, this can be repared by introducing
a quantization map $\rho$, whose path integral interpretation is not
clear at the moment, but is related to the subtelties of computing path
integrals in the presence of a boundary, as we proceed to discuss.
\subsection{Discussion}
The above treatment of the path integral has some subtelties due to
the presence of a boundary and boundary observables. 
They are not fully understood yet. We expect that a more accurate
treatment of these subtelties will give a formula in terms of
Feynman diagrams for a quantization map $\rho$ and thus for a
star product on a general Poisson manifold. Such a description would be more
explicit than the recursive construction described in the previous sections.

\subsubsection{Global gauge fixing}
One might think that our definition of the antighost and Lagrange multiplier
has been too sloppy.
In fact, one should 
have defined them {\em before}\/
choosing the particular point $x$. So let us proceed this way.
Denote by $\gamma$ and $\lambda$
in $\Gamma(\Sigma,X^*TM)$ the ``correct'' antighost
and Lagrange multiplier. We can relate them to $\bar c$ and $\mu$
by
\begin{equation}\label{gammac}
\gamma=\dd\phi_x(B)\bar c,\qquad
\lambda=\dd\phi_x(B)\mu,\qquad
\gamma^+=(\dd\phi_x(B))^{-1,T}\bar c^+,\qquad
\lambda^+=(\dd\phi_x(B))^{-1,T}\mu^+.
\end{equation}
Now the problem is that the transformations \eqref{sfphix}, \eqref{sfphiB} 
and \eqref{gammac} do not provide a canonical transformation
$(\sfB,\sfA,\bar c,\mu,\bar c^+,\mu^+)\rightsquigarrow(\sfeta,\sfX,\gamma,\lambda,\gamma^+,\lambda^+)$.
One may correct for this as proposed in \citen{BLN}; viz., replacing 
\eqref{sfphiB} by
\[
\sfeta_i(u) = ((\dd\phi_{x}(\sfB(u)))^{-1,T}\,\sfA(u))_i
-\phiy ji T^r_{js}(B(u))(\mu^{+}_r\mu^s+\bar c^{+}_r\bar c^s),
\qquad
\forall u\in\Sigma
\]
(which actually modifies only the equation containing $X^+$), where
\[
T_{ij}^k := \phiy kr \frac{\de^2\phi^r}{\de y^i\de y^j}.
\]
This means that the correct BV action reads
\[
\sfS =  \int_{\Pi T\Sigma}\left(-\braket{D\sfB}\sfA+
\frac12\alpha_\phi(x;\sfB)(\sfA,\sfA)\right)\;\sfmu
-\int_\Sigma  \mu^i\bar c^+_i +
\alpha_\phi(x;B)^{ij}c_i T^r_{js}(\mu^{+}_r\mu^s+\bar c^{+}_r\bar c^s).
\]
However, after gauge-fixing as above,
the only extra term produced is
\[
\int_\Sigma 
\alpha_\phi(x;B)^{ij}c_i T^r_{js}\dd{*A_r}\bar c^s,
\]
which vanishes because of the gauge fixing condition.

Notice that, as suggested in \citen{BLN},  
one can add extra terms to the action to make it transform better
under target diffeomorphisms. These terms depend on a torsion-free connection
$\Gamma$ (not necessarily the same as the one appearing in $\phi$)
and its curvature $R$ and have the form
\[
\int_\Sigma
\alpha^{ir}(X)\beta_r\Gamma^j_{ik}(X)(\gamma^+_j\gamma^k+\lambda^+_j\lambda^k)
-R^i_{jkl}(X)\alpha^{jr}(X)\alpha^{ks}(X)\beta_r\beta_s\gamma^l\lambda^+_s.
\]
After the field redefinition we still have the sum of three terms:
the first proportional to $\bar c^+\bar c$, the second to $\mu^+\mu$,
and the third to $\bar c\mu^+$. Since our gauge fixing sets $\mu^+$
to zero, only the first term survives. It turns out however
to be proportional to $\dd{*A}$, i.e., to the gauge-fixing condition;
so it also disappears.

In conclusion, treating antighost and Lagrange multiplier correctly 
does not change our discussion if the gauge-fixing is chosen as in
\eqref{Psi}. In particular, this does not change the bullet product of
two functions.

\subsubsection{Boundary observables}
Putting an observable directly on the boundary---like, e.g.,
$\calO_0(f):=f_\phi(x;B(u_0))$, $u_0\in\de\Sigma$---may be too singular. 
One might look instead for an observable $\calO(f)$
that coincides with the above in the 
classical limit but is less singular otherwise. 
One may then define $\rho(f)=\langle\calO(f)\rangle$ and hope
that this provides a quantization map in the sense of 
Sect.~\ref{sec-dqPm}. 
It would be interesting to understand if this is
possible. 

Observe that a 
natural choice for $\calO(f)$
suggested by the Hamiltonian approach in \citen{CF2}
is
\[
\calO(f)=
f_\phi\left(x;B(u)+\int_\gamma \alpha_\phi(x,\sfB)\sfA\right),
\]
where $u$ is a point in the interior of $\Sigma$ 
and $\gamma$ is any path connecting $u_0$ to $u$.


\subsubsection{Field redefinition with boundary}
The transformations \eqref{sfphix} and \eqref{sfphiB} provide
a BV-canonical, unimodular map if $\Sigma$ has no boundary.
It is conceivable that a correct
treatment of the boundary (in the presence of boundary observables)
will show that the field redefinition adds to the
action a boundary term
$S_{\rm bry}$---as
suggested, e.g., in \citen{CF3} for the 
treatment of infinitesimal
target diffeomorphisms. One may then define $\rho(f)$ as the expectation
value of $\calO_0(f)$---or 
$\calO(f)$---times $\exp(\ii S_{\rm bry}/\hbar)$.
This $\rho$ should provide a quantization map.

If this program can be achieved, one would get a universal explicit formula
for a quantization map.


\section*{Acknowledgements}
A.~S.~C.~thanks Y.~Maeda for an invitation to the
``International Workshop on Noncommutative Geometry and String Theory,''
at Keio University, March 16--22, 2001. A.~S.~C.~also acknowledges
useful discussions with R.~Bott, M.~Cahen, S.~Gutt, J.~Rawnsley
and A.~Weinstein. G.~F.~is grateful to L.~Tomassini for discussions
and to N.~Nekrasov for explanations on \citen{BLN}.

\end{document}